\documentclass[fleqn]{article}
\usepackage{fleqn}
\usepackage{graphicx}
\usepackage{epsfig}
\usepackage{amsmath}
\usepackage{amssymb}
\usepackage{calc}
\newcommand{\beq}{\begin{equation}}
\newcommand{\eeq}{\end{equation}}
\newcommand{\bea}{\begin{eqnarray}}
\newcommand{\eea}{\end{eqnarray}}
\begin{document}
\Large
\begin{center}
{\bf Black Hole Entropy  and Finite Geometry
}
\end{center}
\large
\vspace*{-.1cm}
\begin{center}
P\'eter L\'evay,$^{1}$ Metod Saniga$^{2}$, P\'eter Vrana$^{1}$ and Petr Pracna$^{3}$
\end{center}
\vspace*{-.4cm} \normalsize
\begin{center}
$^{1}$Department of Theoretical Physics, Institute of Physics,
Budapest University of\\ Technology and Economics, H-1521
Budapest, Hungary \vspace*{.2cm}

$^{2}$Astronomical Institute, Slovak Academy of Sciences\\
SK-05960 Tatransk\' a Lomnica, Slovak Republic

\vspace*{.0cm} and

\vspace*{.0cm}

$^{3}$J. Heyrovsk\' y Institute of Physical Chemistry, Academy of Sciences of the Czech Republic\\
Dolej\v skova 3, CZ-182 23 Prague 8, Czech Republic\\

\vspace*{.3cm} (9 March 2009)

\end{center}

\vspace*{-.3cm} \noindent \hrulefill

\vspace*{.1cm} \noindent {\bf Abstract:}
It is shown that the $E_{6(6)}$ symmetric entropy formula  describing black holes and black strings in $D=5$ is intimately tied to the
geometry of the generalized quadrangle GQ$(2,4)$ with automorphism group the Weyl group $W(E_6)$. The $27$ charges  correspond to the
points and the $45$ terms in the entropy formula to the lines of GQ$(2,4)$. Different truncations with $15, 11$ and $9$ charges are
represented by three distinguished subconfigurations of GQ$(2,4)$, well-known to finite geometers; these are the ``doily" (i.\,e.
GQ$(2,2)$) with $15$, the ``perp-set" of a point with $11$, and the ``grid" (i.\,e. GQ$(2,1)$) with $9$ points, respectively.  In order
to obtain the correct signs for the terms in the entropy formula, we  use a non-commutative labelling for the points of GQ$(2,4)$.
For the $40$ different possible truncations with $9$ charges this labelling yields $120$ Mermin squares --- objects well-known from
studies concerning Bell-Kochen-Specker-like theorems. These results are connected to our previous ones obtained for the $E_{7(7)}$
symmetric entropy formula in $D=4$ by observing that the structure of GQ$(2,4)$ is linked to a particular kind of geometric hyperplane
of the split Cayley hexagon of order two, featuring $27$ points located on $9$ pairwise disjoint lines (a distance-3-spread). We
conjecture that the different possibilities of describing the $D=5$ entropy formula using Jordan algebras, qubits and/or qutrits
correspond  to employing different coordinates for an underlying non-commutative geometric structure based on GQ$(2,4)$.

\vspace*{-.2cm}
\noindent \hrulefill

\section{Introduction}

Recently striking multiple relations have been established between the physics of stringy black hole solutions and quantum information
theory \cite{Duff1,Linde,Lev1,Ferr2,Lev2,Lev3,Ferr3,Duff2,Leron1,Duff3,Leron2,Lev4}. Though this ``black hole analogy" still begs for a firm physical
basis, the underlying correspondences have repeatedly proved to be useful for obtaining new insights into one of these fields by
exploiting the methods established within the other. The main unifying theme in these papers is the correspondence between the
Bekenstein-Hawking entropy formula \cite{B,H} for black-hole and black-string solutions in $D=4$ and $D=5$ supergravities arising from
string/M-theory compactifications and certain entanglement invariants of multi-qubit/-qutrit systems. As a new unifying agent in some
of these papers \cite{Lev2,Lev4} the role of discrete geometric ideas have been emphasized. In particular it has been shown
\cite{Ferr2,Lev2} that the Fano plane with seven points and seven lines with points conveniently labelled by seven three-qubit states can
be used to describe the structure of the $E_{7(7)}$ symmetric black hole entropy formula of $N=8$, $D=4$ supergravity. Moreover, this geometric
representation based on the fundamental  $56$  dimensional representation of    $E_{7(7)}$ in terms of $28$ electric and $28$ magnetic charges
enabled a diagrammatic understanding of the consistent truncations with $32, 24$ and $8$ charges as a restriction to quadrangles, lines
and points of the Fano plane \cite{Lev2}. Though the Fano plane turned out to be a crucial ingredient also in later studies, this geometric
representation based on the tripartite entanglement of seven qubits has a number of shortcomings \cite{Duff3}. In order to eliminate these,
in our latest paper \cite{Lev4} we attempted to construct a new representation using merely {\it three-qubits}. The basic idea was to use
the central quotient of the three-qubit Pauli group \cite{Nielsen}, well-known from studies concerning quantum error correcting codes.
This Abelian group can be described by the $63$ {\it real} operators of the Pauli group with multiplication up to a sign. These $63$
operators can be mapped bijectively to the points of a finite geometrical object called the split Cayley hexagon of order two having
$63$ points and $63$ lines, with a subgeometry (the complement of one of its geometric hyperplanes) being the Coxeter graph with $28$
points/vertices. This graph has been related to the charge configurations of the $E_7$ symmetric black hole entropy formula \cite{Lev4}.
The advantage of this representation was a clear understanding of an automorphism of order seven relating the {\it seven} STU subsectors
of $N=8$, $D=4$ spergravity and the explicit appearance of a discrete $PSL(2,7)$ symmetry of the black hole entropy formula.
The permutation symmetry of the STU model (triality) in this picture arises as a subgroup of $PSL(2,7)$.

Encouraged by the partial success of finite geometric ideas in the $D=4$ case the aim of the present paper is to shed some light on a
beautiful finite geometric structure underlying also the $E_{6(6)}$ symmetric entropy formula in $D=5$. 
We show that in this case the relevant
finite geometric objects are {\it generalized quadrangles} with lines of size three.
As it is well-known
black holes in $D=5$ have already played a special role in string theory, since these objects provided the first clue how to understand the microscopic origin of the Bekenstein-Hawking entropy \cite{Vafa}.

As a first step, in Section 2 we emphasize that according to several well-known theorems \cite{feit-higman,bose-shrikhande,higman} we have
just {\it four} (including also a ``weak/degenerate" one made of all lines passing through a fixed point \cite{brouwer-haemers}) such
quadrangles, which are directly related to the {\it four} possible division algebras. It is well-known that magic $N=2$, $D=5$
supergravities \cite{Gun1,Gun2,Gun3,Duff3} coupled to $5,8,14$ and $26$ vector multiplets with symmetries $SL(3,{\bf R})$, $SL(3,{\bf C})$,
$SU^{\ast}(6)$ and $E_{6(-26})$ can be described by Jordan algebras of $3 \times 3$ Hermitian matrices with entries taken from the real
and complex numbers, quaternions and octonions. It is also known that in these cases we have black hole solutions that have cubic
invariants whose square roots yield the corresponding black hole entropy \cite{Gimon}. Moreover, we can also replace in these Jordan
algebras the division algebras  by their split versions. For example, in this way in the case of split octonions we arrive at the $N=8$,
$D=5$ supergravity \cite{Maldacena} with $27$ Abelian gauge fields transforming in the fundamental of $E_{6(6)}$. In this theory the
corresponding black hole solutions have an entropy formula having $E_{6(6)}({\bf Z})$ symmetry \cite{FeKa,Laura,Duff3}. This analogy
existing between division algebras, Jordan algebras and generalized quadrangles with lines having three points leads us to a conjecture
that such finite geometric objects should be relevant for a fuller geometrical understanding of black hole entropy in $D=5$.

In Section 3, by establishing an explicit mapping between the $27$ points and $45$ lines of the generalized quadrangle GQ$(2,4)$ and the
$27$ charges and the $45$ terms in the cubic invariant appearing in the entropy formula, we prove that our conjecture is true.
The crucial observation here is that the automorphism group of GQ$(2,4)$ is the Weyl group $W(E_6)$ with  order $51840$.
Our labelling for the points of GQ$(2,4)$ used here is a one directly related to the two qutrit states of Duff and Ferrara \cite{Ferr3}.
By using the vocabulary of Borsten et al. \cite{Duff3}, this labelling directly relates to the usual one featuring cubic Jordan algebras.

In Section 4 we observe that our geometric correspondence merely gives the number and structure of the terms in the cubic invariant. In
the case of the $E_{6(6)}({\bf Z})$ symmetric black hole entropy in order to produce also the correct signs of these terms we have to
employ a noncommutative labelling for the points of GQ$(2,4)$. To link these considerations to our previous paper on the $E_{7(7)}({\bf
Z})$ symmetric black hole entropy in $D=4$, we adopt the labelling by real three-qubit operators of the Pauli group. We show that this
labelling scheme is connected to a certain type of geometric hyperplane of the split Cayley hexagon of order two featuring precisely
$27$ points that lie on $9$ pairwise disjoint lines. There are $28$ different hyperplanes of this kind in the hexagon, giving rise to
further possible labellings. Next, we focus on special subconfigurations of GQ$(2,4)$ which are called grids. These are generalized
quadrangles GQ$(2,1)$, featuring $9$ points and $6$ lines that can be  arranged in the form of squares. There are 120 distinct copies of
them living within GQ$(2,4)$, grouped to $40$ triples such that each of them comprises all of the  $27$ points of GQ$(2,4)$. Our
noncommutative labelling renders these grids to Mermin squares, which are objects of great relevance for obtaining very economical proofs
to Bell-Kochen-Specker-like theorems.
In order to complete the paper, we also present the action of the Weyl group
on the noncommutative labels of GQ($2,4)$.
This also provides a proof for the $W(E_6)$ invariance of the cubic invariant.

Finally, Section 5 highlights our main findings and presents our conclusive remarks and conjectures. In particular, we conjecture that
the different possibilities of describing the $D=5$ entropy formula using Jordan algebras, qubits and/or qutrits correspond  to employing
different coordinates for an underlying noncommutative geometric structure based on GQ$(2,4)$.

\section{Jordan algebras and generalized quadrangles}

\subsection{Cubic Jordan algebras}

As we remarked in the introduction the charge configurations of $D=5$ black holes/strings are related to the structure of cubic Jordan
algebras. An element of a cubic Jordan algebra can be represented as a $3\times 3$ Hermitian matrix with entries taken from a division
algebra ${\bf A}$, i.\,e. ${\bf R}$, ${\bf C}$, ${\bf H}$ or ${\bf O}$. (The real and complex numbers, the quaternions and the octonions.)
Explicitly, we have

\beq
J_3(Q)=\begin{pmatrix}q_1&Q^v&\overline{Q^s}\\ \overline{Q^v}&q_2&Q^c\\ Q^s&\overline{Q^c}&q_3\end{pmatrix}\qquad q_i\in
{\bf R},\qquad Q^{v,s,c}\in {\bf A},
\eeq
\noindent
where an overbar refers to conjugation in ${\bf A}$. These charge configurations describe electric black holes of the $N=2$, $D=5$
magic supergravities \cite{Gun1,Gun2,Gun3,Duff3}. In the octonionic case the superscripts of $Q$ refer to the fact that the fundamental
$27$ dimensional representation of the U-duality group $E_{6(-26)}$ decomposes under the subgroup $SO(8)$ to three $8$ dimensional
representations (vector, spinor and conjugate spinor) connected by triality and to three singlets corresponding to the $q_i, i=1,2,3$.
Note that a general element in this case is of the form $Q=Q_0+Q_1e_1+\dots +Q_7e_7$, where the ``imaginary units" $e_1,e_2,\dots,e_7$
satisfy the rules of the octonionic multiplication table \cite{Duff3}. The norm of an octonion is $Q\overline{Q}=(Q_0)^2+ \dots +(Q_7)^2$.
The real part of an octonion is defined as ${\rm Re}(Q)=\frac{1}{2}(Q+\overline{Q})$.

The magnetic analogue of $J_3(Q)$ is

\beq
J_3(P)=\begin{pmatrix}p^1&P^v&\overline{P^s}\\ \overline{P^v}&p^2&P^c\\ P^s&
\overline{P^c}&p^3\end{pmatrix}\qquad p^i\in {\bf R},\qquad P^{v,s,c}\in {\bf A},
\eeq
\noindent
describing black strings related to the previous case by the electric-magnetic duality.
The black hole entropy is given by the cubic invariant
\beq
I_3(Q)=q_1q_2q_3-(q_1Q^s\overline{Q^s}+q_2Q^c\overline{Q^c}+q_3Q^v\overline{Q^v})+2{\rm Re}(Q^vQ^sQ^c),
\label{i3}
\eeq
\noindent
as
\beq
S=\pi\sqrt{I_3(Q)},
\label{entropy}
\eeq
\noindent
and for the black string we get a similar formula with $I_3(Q)$ replaced by $I_3(P)$.
Recall that $I_3$ is just the norm of the cubic Jordan algebra and the norm preserving group is $SL(3,{\bf A})$ and $J_3^{\bf A}$
transforms under this group with respect to the $3{\rm dim}{\bf A}+3$ dimensional representation, i.\,e. as the
${\bf 6}$, ${\bf 9}$, ${\bf 15}$ and ${\bf 27}$ of the groups
$SL(3,{\bf R})$, $SL(3,{\bf C})$, $SU^{\ast}(6)$ and $E_{6(-26)}$.

We can also consider cubic Jordan algebras with ${\bf C}$, ${\bf H}$ and ${\bf O}$ replaced by the corresponding {\it split versions}.
In the octonionic case ${\bf O}_s$ the ``norm" is defined as
\beq
Q\overline{Q}=(Q_0)^2+(Q_1)^2+(Q_2)^2+(Q_3)^2-(Q_4)^2-(Q_5)^2-(Q_6)^2-(Q_7)^2,
\label{splitnorm}
\eeq
\noindent
and the group preserving the norm of the corresponding Jordan algebra is $E_{6(6)}$, which decomposes similarly under $SO(4,4)$.
This is the case of $N=8$ supergravity with duality group $E_{6(6)}$ \cite{Gun4}.
Note that the groups $E_{6(-26)}$ and $E_{6(6)}$ are the symmetry groups of the corresponding classical supergravity. In the quantum
theory the black hole/string charges become integer-valued and the relevant $3 \times 3$ matrices are defined over the {\it integral}
octonions and {\it integral} split octonions, respectively. Hence, the U-duality groups are in this case broken to $E_{6(-26)}({\bf Z})$
and $E_{6(6)}({\bf Z})$ accordingly. In all these cases the entropy formula is given by Eqs.\,(\ref{i3})--(\ref{entropy}),
with the norm given by either the usual one or its split analogue, Eq. (\ref{splitnorm}).

It is also important to recall that the magic $N=2$ supergravities associated with the real and complex numbers and the quaternions can be
obtained as consistent reductions of the $N=8$ one \cite{Gimon} which is based on the {\it split octonions}. On the other hand, the $N=2$
supergravity based on the division algebra of the octonions is exceptional since it is the only one that cannot be obtained from the split
octonionic $N=8$ one by truncation.

\subsection{Finite generalized quadrangles}

Now we summarize the basic definitions on generalized quadrangles that we'll need later.
A {\it finite generalized quadrangle} of order $(s, t)$, usually denoted GQ($s, t$), is an incidence structure $S = (P, B, {\rm I})$,
where $P$ and $B$ are disjoint (non-empty) sets of objects, called respectively points and lines, and where I is a symmetric point-line
incidence relation satisfying the following axioms \cite{paythas}: (i) each point is incident with $1 + t$ lines ($t \geq 1$) and two
distinct points are incident with at most one line; (ii) each line is incident with $1 + s$ points ($s \geq 1$) and two distinct lines
are incident with at most one point;  and (iii) if $x$ is a point and $L$ is a line not incident with $x$, then there exists a unique
pair $(y, M) \in  P \times B$ for which $x {\rm I} M {\rm I} y {\rm I} L$; from these axioms it readily follows that $|P| = (s+1)(st+1)$
and $|B| = (t+1)(st+1)$. It is obvious that there exists a point-line duality with respect to which each of the axioms is self-dual.
Interchanging points and lines in $S$ thus yields a generalized quadrangle $S^{D}$ of order $(t, s)$, called the dual of $S$. If $s = t$,
$S$ is said to have order $s$. The generalized quadrangle of order $(s, 1)$ is called a grid and that of order $(1, t)$ a dual grid. A
generalized quadrangle with both $s > 1$ and $t > 1$ is called thick. In any GQ$(s,t)$, $s+t$ divides both $st(1+st)$ \cite{feit-higman}
and $st(s+1)(t+1)$ \cite{bose-shrikhande}; moreover, if $s > 1$ (dually, $t > 1$) then $t \leq s^2$ (dually, $s \leq t^2$) \cite{higman}.

Given two points $x$ and $y$ of $S$ one writes $x \sim y$ and says that $x$ and
$y$ are collinear if there exists a line $L$ of $S$
incident with both. For any $x \in P$ denote                                    $x^{\perp} = \{y \in P | y \sim x \}$
 and note that $x \in x^{\perp}$;  obviously, $x^{\perp}
= 1+s+st$. Given an arbitrary subset $A$ of $P$, the {\it perp}(-set) of $A$, $A
^{\perp}$, is defined as $A^{\perp} = \bigcap \{x^{\perp}
| x \in A\}$ and $A^{\perp \perp} := (A^{\perp})^{\perp}$.
An ovoid of a generalized
 quadrangle $S$ is a set of points of $S$ such that each
 line of $S$ is incident with exactly one point of the set;  hence, each ovoid
contains $st + 1$ points.

A {\it geometric hyperplane} $H$ of a point-line geometry $\Gamma (P,B)$ is a   proper
subset of $P$ such that each line of $\Gamma$ meets
$H$ in one or all points \cite{ron}. For $\Gamma =$ GQ($s, t$), it is well known
 that $H$ is one of the following three kinds: (i) the
 perp-set of a point $x$,  $x^{\perp}$; (ii) a (full) subquadrangle of order ($s,
 t'$), $t' < t$; and (iii) an ovoid.

In what follows, we shall be uniquely concerned with generalized
quadrangles having lines of size {\it three}, GQ$(2,t)$. From the
above-given restrictions on parameters of GQ$(s,t)$ one readily
sees that these are of three distinct kinds, namely GQ$(2,1)$,
GQ$(2,2)$ and GQ$(2,4)$, each unique. They can uniformly be
characterized as being formed by the points and lines of a
hyperbolic, a parabolic and an elliptic quadric in three-, four-
and five-dimensional projective space over GF(2), respectively.
GQ$(2,1)$ is a grid of 9 points on 6 lines, being the complement
of the lattice graph $K_3 \times K_3$. It contains only ovoids (6;
each of size 3) and perp-sets (9; each of size 5). GQ$(2,1)$  is
obviously different from its dual, the complete bipartite graph on
6 vertices. GQ$(2,2)$ is the smallest thick generalized
quadrangle, also known as the ``doily." This quadrangle is endowed
with 15 points/lines, with each line containing 3 points and,
dually, each point being on 3 lines; moreover, it is a self-dual
object, i.\,e., isomorphic to its dual. It is the complement of
the triangular graph $T(6)$ and features all the three kinds of
geometric hyperplanes, of the following cardinalities: 15
perp-sets, $x^{\perp}$, 7 points each; 10 grids (i.\,e.
GQ$(2,1)$s), 9 points each; and 6 ovoids, 5 points each. One of
its most familiar constructions is in terms of synthemes and
duads, where the point set consists of all pairs of a six-element
set and the line set comprises all three-sets of pairs forming a
partition of  the six-element set. The full group of automorphisms
of GQ$(2,2)$ is $S_6$, of order 720. The last case in the
hierarchy is GQ$(2,4)$, which possesses 27 points and 45 lines,
with lines of size 3 and 5 lines through a point. Its full group
of automorphisms is of order 51840, being isomorphic to the Weyl
group $W(E_6)$. GQ$(2,4)$ is obviously not a self-dual structure;
its dual, GQ(4,2), features 45 points and 27 lines, with lines of
size 5 and 3 lines through a point. Unlike its dual, which
exhibits all the three kinds of geometric hyperplanes, GQ(2,4) is
endowed only with perp-sets (27, of cardinality 11 each) and
GQ$(2,2)$s (36), {\it not} admitting any ovoid. One of its
constructions goes as follows. One starts with the
above-introduced syntheme-duad construction of GQ$(2,2)$, adds 12
more points labelled simply  as $1,2,3,4,5,6,1',2',3',4',5',6'$
and defines 30 additional lines as the three-sets
$\{a,b',\{a,b\}\}$ of points, where $a,b \in \{1,2,3,4,5,6\}$ and
$a \neq b$ --- as diagrammatically illustrated, after Polster
\cite{polster},  in Figure 1. To conclude this section, we
emphasize the fact that GQ$(2,1)$ is a geometric hyperplane of
GQ$(2,2)$, which itself is a geometric hyperplane of GQ$(2,4)$.
\begin{figure}[t]
\centerline{\includegraphics[width=7.5truecm,clip=]{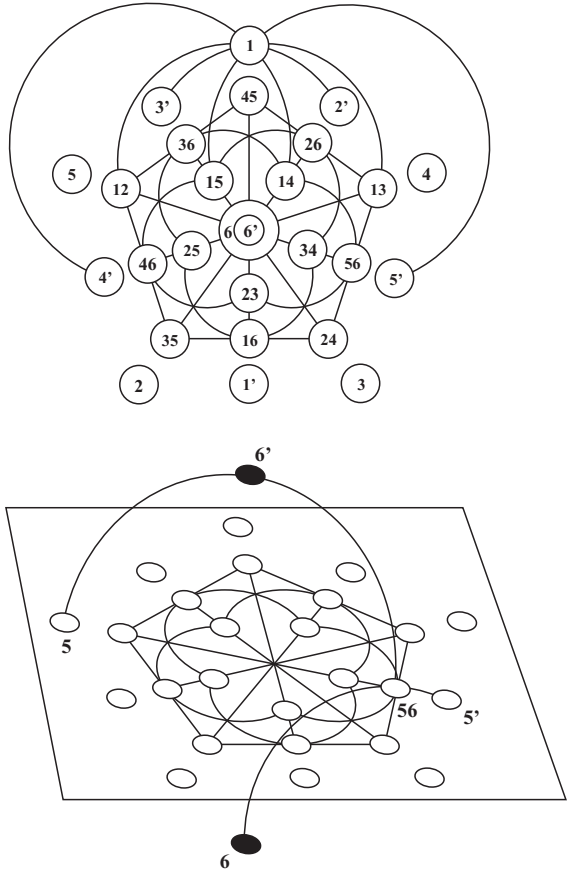}}
\vspace*{0.5cm} \caption{A diagrammatic illustration of the
structure of the generalized quadrangle GQ($2,4)$ after Polster
\cite{polster}. In both the figures, each picture depicts all 27
points (circles). The top picture shows only 19 lines (line
segments and arcs of circles) of GQ$(2,4)$, with the two points
located in the middle of the doily being regarded as lying one
above and the other below the plane the doily is drawn in. 16 out
of the missing 26 lines can be obtained by successive rotations of
the figure through 72 degrees around the center of the pentagon.
The bottom picture shows a couple of lines which go off the
doily's plane; the remaining 8 lines of this kind are again got by rotating the
figure through 72 degrees around the center of the pentagon.}
\end{figure}

Looking at the sequence of numbers $27$, $15$ and $9$, representing the number of points of these quadrangles, one is immediately tempted
to relate these numbers to the dimensions of the representations of the norm-preserving groups of the cubic Jordan algebras based on
${\bf O}$ (or ${\bf O}_s$), ${\bf H}$ and ${\bf C}$. After a quick glance at the structure of the corresponding entropy formulas
\cite{Gimon} constructed within the context of magic supergravities one also recognizes that the sequence $45$, $15$ and $6$, representing
the number of lines, should correspond to the number of different terms in the corresponding entropy formulas. (Thus, for example, $15$ is
the number of terms in the Pfaffian of a $6 \times 6$ antisymmetric matrix giving rise to the quaternionic magic entropy formula. Moreover,
$6$ is the number of terms in the determinant of a $3 \times 3$ matrix occurring in the entropy formula of the complex case.) Indeed,
using the nice labelling scheme developed by Duff and his coworkers\cite{Duff3} it is not difficult to set up an explicit geometric correspondence
between the $45$ lines of GQ$(2,4)$ and the terms in the cubic invariant of Eq. (\ref{i3}). This is the task we now turn to.
\section{The cubic invariant and GQs with lines of size three}

\subsection{GQ$(2,4)$ and qutrits}
Since except for the octonionic magic all the $N=2$ magic supergravities can be obtained as consistent truncations of the $N=8$
split-octonionic case, let us consider the cubic invariant $I_3$ of Eq.\,(\ref{i3}) with the $U$-duality group $E_{6(6)}$. Let us
consider the decomposition of the $27$ dimensional fundamental representation of $E_{6(6)}$ with respect to its $SL(3,{\bf R})^{\otimes
3}$ subgroup.
We have the decomposition
\beq
E_{6(6)}\supset SL(3,{\bf R})_A\times SL(3,{\bf R})_B\times SL(3,{\bf R})_C
\label{griddecomp}
\eeq
\noindent
under which
\beq
{\bf 27}\to ({\bf 3}^{\prime},{\bf 3},{\bf 1})\otimes ({\bf 1},{\bf 3}^{\prime},{\bf 3}^{\prime})\otimes ({\bf 3},{\bf 1},{\bf 3}).
\eeq
\noindent
As it is known \cite{Ferr3,Duff3}, the above-given decomposition is related to the ``bipartite entanglement of three-qutrits"
interpretation of the ${\bf 27}$ of $E_6({\bf C})$. Neglecting the details, all we need is three $3 \times 3$ real matrices $a,b$
and $c$ with the index structure
\beq
{a^A}_B,\qquad {b^{BC}},\qquad {c_{CA}},\qquad  A,B,C=0,1,2,
\label{ampl}
\eeq
\noindent
where the upper indices are transformed according to the (contragredient) ${\bf 3}^{\prime}$
and the lower ones by ${\bf 3}$.
Then according to the dictionary developed in Borsten et al. \cite{Duff3},
we have
\beq
p^1=-{a^0}_0,\qquad p^2=-{a^1}_1,\qquad p^3=-{a^3}_3,
\label{0v}
\eeq
\noindent
\begin{eqnarray}
2P^c=&-&({a^1}_2+{a^2}_1)e_0-(b^{00}+c_{00})e_1-(b^{01}+c_{10})e_2
-(b^{02}+c_{20})e_3\nonumber\\
&+&({a^1}_2-{a^2}_1)e_4+(b^{00}-c_{00})e_5+(b^{01}-c_{10})e_6+(b^{02}-c_{20})e_7,
\label{1}
\end{eqnarray}
\noindent
\begin{eqnarray}
2P^s=&-&({a^2}_0+{a^0}_2)e_0-(b^{10}+c_{01})e_1-(b^{11}+c_{11})e_2
-(b^{12}+c_{21})e_3\nonumber\\
&+&({a^2}_0-{a^0}_2)e_4+(b^{10}-c_{01})e_5+(b^{11}-c_{11})e_6+(b^{12}-c_{21})e_7
,
\label{2}
\end{eqnarray}
\noindent
\begin{eqnarray}
2P^v=&-&({a^0}_1+{a^1}_0)e_0-(b^{20}+c_{02})e_1-(b^{21}+c_{12})e_2
-(b^{22}+c_{22})e_3\nonumber\\
&+&({a^0}_1-{a^1}_0)e_4+(b^{20}-c_{02})e_5+(b^{21}-c_{12})e_6+(b^{22}-c_{22})e_7.
\label{3}
\end{eqnarray}
We can express $I_3$ of Eq.\,(\ref{i3}) in the alternative form
as
\beq
I_3={\rm Det}J_3(P)=a^3+b^3+c^3+6abc.
\label{i3v}
\eeq
\noindent
Here
\beq
a^3=\frac{1}{6}{\varepsilon}_{A_1A_2A_3}{\varepsilon}^{B_1B_2B_3}{a^{A_1}}_{B_1}
{a^{A_2}}_{B_2}{a^{A_3}}_{B_3},
\eeq
\noindent
\beq
b^3=\frac{1}{6}{\varepsilon}_{B_1B_2B_3}{\varepsilon}_{C_1C_2C_3}
b^{B_1C_1}b^{B_2C_2}b^{B_3C_3},
\eeq
\noindent
\beq
c^3=\frac{1}{6}{\varepsilon}^{C_1C_2C_3}{\varepsilon}^{A_1A_2A_3}c_{C_1A_1}
c_{C_2A_2}c_{C_3A_3},
\eeq
\noindent
\beq
abc=\frac{1}{6}{a^{A}}_Bb^{BC}c_{CA}.
\eeq
\noindent
Notice that the terms like $c^3$ produce just the determinant of the corresponding $3 \times 3$ matrix. Since each determinant contributes
$6$ terms, altogether we have $18$ terms from the first three terms in Eq.\,(\ref{i3v}). Moreover, since it is easy to see that the fourth
term contains $27$ terms, altogether $I_3$ contains precisely $45$ terms, i.\,e. the number which is equal to that of lines in GQ$(2,4)$.

In order to set up a bijection between the points of GQ$(2,4)$ and the $27$ amplitudes of the two-qutrit states of Eq.\,(\ref{ampl}),
we use the basic ideas of the above-given construction of GQ$(2,4)$ (see Figure 1).
Since the automorphism group of the doily (GQ$(2,2)$) is the symmetric group $S_6$, this construction is based on labelling the $15$
points of the doily by the $15$ two-element subsets of the set $\{1,2,3,4,5,6\}$ on which $S_6$ acts naturally. The next step consists
of adding two six element sets: the basic set $\{1,2,3,4,5,6\}$ and an extra one
$\{1^{\prime},2^{\prime},3^{\prime},4^{\prime},5^{\prime},6^{\prime}\}$ according to the rule as explained in Figure 1.
Hence, the duad labelling is: $(ij), i<j$, $(i)$ and $(j^{\prime})$ where $i,j=1,2,\dots,6$.

We can easily relate the labelling of these $27$ points to the structure of two $8 \times 8$ antisymmetric matrices with $28$ independent
components, each with one special component removed. Let us label the rows and columns of such a matrix by the letters $I,J=0,1,2,\dots,7$.
(The reason for this unusual labelling will be clarified in the next section.) Let us choose the special component to be removed from both
of our $8 \times 8$ matrices to be the element $01$. Then we choose a $6 \times 6$ antisymmetric block in one of the $8 \times 8$ matrices
labelled by the row and column indices $ I<J,\quad I,J=2,3,\dots,7$. Its $15$ components give rise to a duad labelling of the doily. Now,
from the other $8 \times 8$ matrix we choose the elements of the form $0J$ with $J=2,\dots,7$ to correspond to the set
$\{1^{\prime},\dots,6^{\prime}\}$, and the ones of the form $1J$ to the one $\{1,\dots,6\}$. (Clearly, the row and column indices are
shifted by one unit with respect to the usual duad indices, i.\,e. $I=i+1, J=j+1$.) Now, it is well-known that the cubic $E_{6(6)}$
invariant for $D=5$ black hole solutions is related to the quartic $E_{7(7)}$ invariant for $D=4$ ones by a suitable truncation of the
Freudenthal triple system to the corresponding cubic Jordan algebra \cite{Pioline}. Recently, the Freudenthal triple description of the
$D=4$ black hole entropy was related to the usual description due to Cartan using two $8 \times 8$ antisymmetric matrices \cite{Duff3},
corresponding to the $28$ electric and $28$ magnetic charges. Using Table 32 of Ref. 10, giving a dictionary between these descriptions,
it is easy to realize that the $27$ elements of the cubic Jordan algebra $J_3(P)$ split as $27 = 15 + 12$ between these two $8 \times 8$
matrices. This automatically defines a one-to-one mapping between the duad construction of GQ$(2,4)$ and the $27$ elements of $J_3(P)$.
As the last step, using Eqs.\,(\ref{0v}--\ref{3}) we can readily relate the arising $J_3(P)$ labelling of GQ$(2,4)$ to the one in terms of
two-qutrit amplitudes of Eq.\,(\ref{ampl}).
The explicit relationship between the duad labelling and the qutrit one is as follows
\beq
\{1,2,3,4,5,6\}=\{c_{21},{a^2}_1,b^{01},{a^0}_1,c_{01},b^{21}\},
\eeq
\noindent
\beq
\{1^{\prime},2^{\prime},3^{\prime},4^{\prime},5^{\prime},6^{\prime}\}=
\{b^{10},c_{10},{a^1}_2,c_{12},b^{12},{a^1}_0\},
\eeq
\noindent
\beq
\{12,13,14,15,16,23,24,25,26\}=\{c_{02},b^{22},c_{00},{a^1}_1,b^{02},{a^0}_0,
b^{11},c_{22},{a^0}_2\},
\eeq
\noindent
\beq
\{34,35,36,45,46,56\}=\{{a^2}_0,b^{20},c_{11},c_{20},{a^2}_2,b^{00}\}.
\eeq
\noindent
This relationship is easily grasped by comparing Figure 2, which depicts the qutrit labelling, with Figure 1 (top).

Next, notice that the lines of GQ$(2,4)$ are of two types. They are either of the form $(i,ij,j^{\prime})$ or  $(ij,kl,mn)$, where
$i,j,j^{\prime},\dots=1,\dots,6$ and $i,j,k,l,m,n$ are different. We have $30$ lines of the first and $15$ lines of the second type.
The latter ones belong to the doily.
Notice also that the three two-qutrit states of Eq.\,(\ref{ampl}) partition the $27$ points of GQ$(2,4)$ to $3$
disjoint grids, i.\,e. GQ$(2,1)$s. The points of these three grids are coloured differently (in an online version only).
The $27$ lines corresponding to the terms of ${\rm Tr}(abc)$ of Eq.\,(\ref{i3v})
are of the type like the one ${a^1}_2b^{22}c_{21}$, and the $3\times 6=18$ terms
are coming from the three $3\times 3$ determinants $a^3,b^3,c^3$. These terms are of the form like the one $b^{20}b^{02}b^{11}$.
From Figure 2 one can check that each of $45$ lines of GQ$(2,4)$ corresponds to exactly one monomial of Eq.\,(\ref{i3v}).

We close this subsection with an important comment/observation. It is well-known that the automorphism group of the generalized quadrangle
GQ$(2,4)$ is the Weyl group\cite{paythas}  $W(E_6)$ of order 51840. Moreover, the cubic invariant is  also connected to the geometry of smooth
(non-singular)) cubic  surfaces in ${\bf CP}^3$. It is a classical result that the automorphism group of the configuration of $27$
lines on a cubic can also be identified with $W(E_6)$. It is also known that different configurations of lines are related to special
models of exceptional Lie algebras \cite{Manivel}. Indeed, it was Elie Cartan who first realized \cite{Cartan} that the $45$ monomials of
our cubic form stabilized by $E_6$ are in correspondence with the tritangent planes of the qubic.
In the light of this fact, our success in parametrizing the monomials of $I_3$ using the lines of GQ$(2,4)$ is not at all surprising.
\begin{figure}[t]
\centerline{\includegraphics[width=10truecm,clip=]{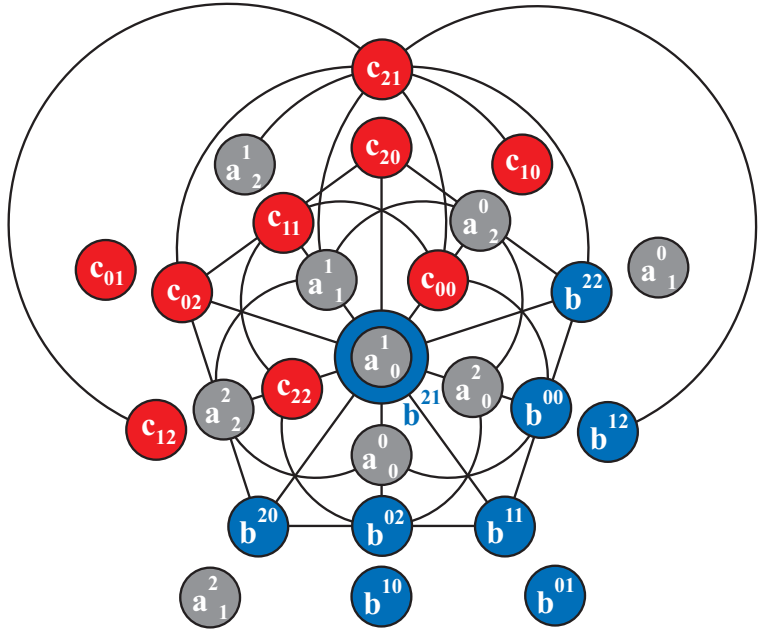}}
\vspace*{0.5cm} \caption{A qutrit labelling of the points of GQ$(2,4)$. Three different colours (online only) are used to
illustrate a triple of grids partitioning the point set.}
\end{figure}
\subsection{Geometric hyperplanes and truncations}

Let us focus now on geometric hyperplanes of GQ$(2,4)$. As already
mentioned in Sec.\,2.2, the only type of hyperplanes featured by
GQ$(2,4)$ are doilies (36) and  perp sets (27). Moreover,
GQ$(2,4)$ also contains $3 \times 40 = 120$ grids; however, these
are {\it not} its geometric hyperplanes \cite{SLPV}. (This is
quite different from the GQ$(2,2)$ case, where grids are geometric
hyperplanes.) Though they are not hyperplanes, they have an
important property that there exits $40$ triples of them, each
partitioning the point set of GQ$(2,4)$.

It is easy to find a physical interpretation of the hyperplanes of GQ$(2,4)$. The doily has $15$ lines, hence we should have a truncation
of our cubic invariant which has $15$ charges. Of course, we can interpret this truncation in many different ways corresponding to the
$36$ different doilies residing in our GQ$(2,4)$. One possibility is a truncation related to the one which employs instead of
the split octonions, the split quaternions in our $J_3(P)$.
The other is to use ordinary quaternions inside our split octonions, yielding
the Jordan algebras corresponding to the quaternionic magic.
In all these cases the relevant entropy formula is related to the Pfaffian of an antisymmetric $6 \times 6$ matrix ${\cal A}^{jk},
\quad i,j=1,2,\dots,6$, defined as
\beq
{\rm Pf}(A)\equiv\frac{1}{3!2^3}{\varepsilon}_{ijklmn}A^{ij}A^{kl}A^{mn}.
\label{Pfaff}
\eeq
\noindent
The simplest way of finding a decomposition of $E_{6(6)}$ directly related to a doily sitting inside GQ$(2,4)$ is the
following one \cite{Laura2,Duff3,Baez}
\beq
E_{6(6)}\supset SL(2)\times SL(6)
\eeq
\noindent
under which
\beq
{\bf 27}\rightarrow ({\bf 2},{\bf 6})\oplus ({\bf 1},{\bf 15}).
\eeq
\noindent
Clearly, this decomposition is displaying nicely its connection with the duad construction of GQ$(2,4)$.
One can show that under this decomposition $I_3$ schematically factors as
\beq
I_3={\rm Pf}(A)+u^TAv,
\label{Pfaffdecomp}
\eeq
\noindent
where $u$ and $v$ are two six-component vectors.
We will have something more to say about this decomposition in the next section.

The next important type of subconfiguration of GQ$(2,4)$ is the grid.
As we have already remarked, grids are {\it not} geometric hyperplanes of GQ$(2,4)$.
The decomposition underlying this type of subconfiguration is the one given by Eq.\,(\ref{griddecomp}).
It is also obvious that the $40$ triples of pairwise disjoint grids are intimately connected to
the $40$ different ways we can obtain a qutrit description of $I_3$.
Note that there are $10$ grids which are geometric hyperplanes of a particular copy of the doily of GQ$(2,4)$.
This is related to the fact that the quaternionic magic case with $15$ charges can be truncated to
the complex magic case with $9$ ones.

The second type of hyperplanes we should consider are perp-sets.
As we already know, perp-sets are obtained by selecting an arbitrary point and considering all the points collinear with it.
Since we have five lines through a point, any perp set has $1+10=11$ points.
A decomposition which corresponds to perp-sets is thus of the form \cite{Duff3}
\beq
E_{6(6)}\supset SO(5,5)\times SO(1,1)
\eeq
\noindent
under which
\beq
{\bf 27}\rightarrow {\bf 16}_{\bf 1}\oplus {\bf 10}_{\bf -2}\oplus {\bf 1}_{\bf 4}.
\eeq
\noindent
This is the usual decomposition of the $U$-duality group into the $T$ duality and $S$ duality \cite{Duff3}.
It is interesting to see that the last term (i.\,e. the one corresponding to the fixed/central point in a perp-set)
describes the $NS$ five-brane charge. Notice that we have five lines going through this fixed point of a perp-set.
These correspond to the $T^5$  of the corresponding compactification. The two remaining points on each of these $5$ lines correspond
to $2\times 5=10$ charges.  They correspond to the $5$ directions of $KK$ momentum and the $5$ directions of fundamental string winding.
In this picture the $16$ charges {\it not belonging to} the perp-set correspond to the $16$ D-brane charges.
Notice that we can get $27$ similar truncations based on the $27$ possible central points of the perp-set.
For a group theoretical meaning of the corresponding decomposition of the cubic invariant, see the paper by Borsten et al. \cite{Duff3}.

\section{Noncommutative coordinates for GQ$(2,4)$}

\subsection{GQ$(2,4)$ and qubits}
The careful reader might have noticed that there is one important issue we have not clarified yet.
What happened to the {\it signs} of the terms in the cubic invariant?
Can we account for them via some sort of geometric construction?

In order to start motivating the problem of signs, we observe that
the terms that should contain negative signs are the first three ones of Eq.\,(\ref{i3v}), containing determinants of $3 \times 3$
matrices.
Indeed, the labelling of Figure 2 only produces the terms of the cubic invariant $I_3$ {\it up to a sign}.
One could immediately suggest that we should also include a special distribution
of signs to the points of GQ$(2,4)$.
This would take care of the
negative signs in the first three terms of Eq.\,(\ref{i3v}).

However, it is easy to see that no such distribution of signs exists.
The reason for this is as follows. We have a triple of grids inside our quadrangle
corresponding to the three different two-qutrit states.
Truncation to any of such states (say to the one with amplitudes described by the matrix $c$) yields the cubic invariant
$I_3(c)={\rm Det}(c)$.
The structure of this determinant is encapsulated in the structure of the corresponding grid.
We can try to arrange the $9$ amplitudes in a way that the $3$ plus signs for the determinant should occur along the rows and
the $3$ minus signs along the columns.
But this is impossible since multiplying all of the nine signs  ``row-wise"
yields a plus sign, but ``column-wise" yields a minus one.

Readers familiar with the Bell-Kochen-Specker type theorems ruling out noncontextual hidden variable theories may immediately suggest
that if we have failed to associate signs with the points of the grid, what about
trying to use noncommutative objects instead?
More precisely, we can try to associate objects that are generally noncommuting but that are pairwise commuting along the lines.
This is exactly what is achieved by using Mermin squares \cite{Mermin1,Mermin2,Peres}.
Mermin squares are obtained by assigning pairwise commuting  {\it two-qubit Pauli matrices}
to the lines of the grid in such a way that the naive sign assignment does not work,
but we get the identity operators with the correct  signs by multiplying the operators row- and column-wise.

It is known \cite{Sanigadoily} that $15$ of the two-qubit Pauli operators belonging to the {\it two-qubit} Pauli group \cite{Nielsen}
can be associated to the points of the doily in such a way that we have mutually commuting operators along all of its $15$ lines.
Moreover, this assignment automatically yields Mermin squares for the $10$
grids living inside the doily.
Hence, a natural question to be asked is whether it is possible to use the same trick for GQ$(2,4)$?
A natural extension would be to try to label the $27$ points of GQ$(2,4)$
with a special set from the operators of the {three-qubit} Pauli group.
In our recent paper \cite{Lev4} we have already gained some insight into the structure of the central quotient of this
group and its connection to the $E_7$ symmetric black hole entropy in $D=4$. Hence, we can even be more ambitious
and search for three-qubit labels for GQ$(2,4)$ also describing an embedding of our qubic invariant to the quartic one.
In this way we would also obtain a new insight into the connection between the $D=4$ and $D=5$ cases in finite geometric terms.

In order to show that this program can indeed be carried out, let us define
the {\it real} three qubit Pauli operators by introducing the notation \cite{Lev4}
$X\equiv {\sigma}_1$, $Y=i{\sigma}_2$, and $Z\equiv {\sigma}_3$;
here, ${\sigma}_j, j=1,2,3$ are the usual $2 \times 2$ Pauli matrices.
Then we can define the real operators of the three-qubit Pauli group by forming the tensor products of the form $ABC \equiv
A \otimes B \otimes C$ that are $8 \times 8$ matrices. For example, we have
\beq
ZYX\equiv Z\otimes Y\otimes X=\begin{pmatrix}Y\otimes X&0\\0&-Y\otimes X\end{pmatrix}=\begin{pmatrix}0&X&0&0\\-X&0&0&0\\0&0&0&-X\\0&0&X&0\end{pmatrix}.
\label{example}
\eeq
\noindent
Notice that operators containing an even number of $Y$s are {\it symmetric} and the ones containing an odd number of $Y$s are {\it
antisymmetric}.
Disregarding the identity, $III$, ($I$ is the $2\times 2$ identity matrix) we have $63$ of such operators. We have shown \cite{Lev4}
that they can be mapped bijectively to the $63$ points of the split Cayley hexagon of order two in such a way that its $63$ lines
are formed by three pairwise commuting operators.
These $63$ triples of operators have the property that their product equals $III$
{\it up to a sign}.

It is easy to check that the $35$ symmetric operators form a geometric hyperplane of the hexagon. Its complement is the famous Coxeter
graph, whose vertices are labelled by the $28$ antisymmetric matrices.
It was shown \cite{Lev4} that the automorphism group of both of these subconfigurations is $PSL(2,7)$, having a generator of order seven.
Due to this we can group the $28$ antisymmetric operators to $4$ seven-element sets. One of these sets is

\beq
(g_1,g_2, g_3,g_4,g_5,g_6,g_7
)=(IIY,ZYX,YIX,YZZ,XYX,IYZ,YXZ)
\label{cliff}
\eeq
\noindent
satisfying the relation $\{g_a,g_b\}=-2{\delta}_{ab},\quad a,b=1,2,\dots 7$, i.\,e. these operators form the generators of a
seven-dimensional Clifford algebra.
Notice that these generators, up to some sign conventions and a cyclic permutation, are precisely the ones used by Cremmer and Julia
in their classical paper \cite{Julia}
on $SO(8)$ supergravity. Namely, their generators ${\gamma}^a, a=4,5,\dots,10$ have the form
\beq
\{{\gamma}^4,{\gamma}^5,{\gamma}^6,{\gamma}^7,{\gamma}^8,{\gamma}^9,{\gamma}^{10}\}=\{ZYX,-ZYZ,ZIY,XXY,XYI,-XZY,-YII\}.
\eeq
\noindent
The remaining $21$ antisymmetric operators are of the form $\frac{1}{2}[g_a,g_b],\quad a,b=1,2,\dots,7$, i.\,e. they generate an
$so(7)$ algebra.
One can then form the $8 \times 8$ matrix $-{\Gamma}^{IJ}, I,J=0,1,\dots,7$
whose entries are our $28$ antisymmetric matrices
\beq
-{\Gamma}^{0a}=g_{0a}\equiv g_a,\qquad -{\Gamma}^{ab}=g_{ab}\equiv\frac{1}{2}[g_a,g_b].
\label{so8}
\eeq
\noindent
In other words, $({\Gamma}^{IJ})_{AB}, A,B=0,1,\dots,7$
are generators of the $so(8)$ algebra in the spinor representation.
Hence, we managed to relate the $28$ generators of $so(8)$ to the complement of one of the geometric hyperplanes of the split Cayley
hexagon of order two, namely to the Coxeter graph.

Notice that Eqs. (\ref{cliff}) and (\ref{so8}) give an explicit labelling for the $28$ points of the Coxeter graph in terms of three-qubit
operators.
We can make use of this structure by employing these three-qubit operators for expanding the $N=8$ central charge ${\cal Z}_{AB}$ as
\beq
{\cal Z}_{AB}=-(x^{IJ}+iy_{IJ})({\Gamma}^{IJ})_{AB},
\label{central4}
\eeq
\noindent
where summation for $I<J$ is implied and the real antisymmetric matrices $x^{IJ}$ and $y_{IJ}$  describe the $28$ electric and $28$ magnetic charges which are related to
some numbers of membranes wrapping around the extra dimensions where these objects live in \cite{Becker}.

In order to establish a connection between the $D=4$ and $D=5$ cases, we assign to one of the $28$ antisymmetric three-qubit operators
a special status. Later, we will show that this choice amounts to a choice of the symplectic structure ${\Omega}$ in the usual formalism
of $D=5$ black hole solutions \cite{Gimon}.
Let us make the following choice
\beq
{\Omega}=IIY=g_1=g_{01}=-{\Gamma}^{01}.
\label{omega}
\eeq
\noindent
(The usual choice for ${\Omega}$ is $YII$ \cite{Julia,Gimon}.)
Then recalling the duad construction of $GQ(2,4)$, a natural choice to try for the labelling
of the $27$ points of our quadrangle is
                                                                                \beq                                                                            \{1^{\prime},2^{\prime},3^{\prime},4^{\prime},5^{\prime},6^{\prime}\}\leftrightarrow  \{g_{02},g_{03},g_{04},g_{05},g_{06}, g_{07}\}                           =\{g_2,g_3,g_4,g_5,g_6,g_7\},                                                    \eeq                                                                            \noindent
\beq                                                                            \{1,2,3,4,5,6\}\leftrightarrow\{g_{12},g_{13},g_{14},g_{15},g_{16},g_{17}\},    \eeq                                                                            \noindent
\beq                                                                            \{12,13,14,15,16,23,24,25,26\}\leftrightarrow                                   \{g_{23},g_{24},g_{25},g_{26},g_{27},g_{34},g_{35},g_{36},g_{37}\},             \eeq                                                                            \noindent                                                                       \beq                                                                            \{34,35,36,45,46,56\}\leftrightarrow                                            \{g_{45},g_{46},g_{47},g_{56},g_{57},g_{67}\},                                  \eeq                                                                            \noindent
i.\,e., shifting all the indices of $g_{IJ}$ not containing $0$ or $1$ by $-1$ we get the duad labels.

Now using the explicit form of the antisymmetric operators
$g_{ab}, a,b\neq 0,1$ used to label the points of our generalized
quadrangle we notice that for all of the $45$ lines the product of
the corresponding $3$ three-qubit operators gives, up to a sign,
$\Omega$! Moreover, we also realize that the $15$ triples of
operators associated with the $15$ lines of the doily are pairwise
commuting. However, the triples of operators belonging to the $30$
lines featuring the double-sixes outside the doily fail to be
pairwise commuting. But we also notice that for such lines the $2$
operators belonging to the double-sixes are always commuting, but
either of them anticommutes with the remaining operator belonging
to the doily. It is also clear that ${\Omega}$ {\it anticommutes}
with the operators of the double-sixes, and {\it commutes} with
the ones of the doily. Hence, if we multiply all the operators
belonging to the doily by $\Omega$, the resulting {\it symmetric}
ones will preserve the nice pairwise commuting property, and at
the same time the same property is also achieved for the resulting
{\it antisymmetric} operators featuring the lines of the double
sixes. And as an extra bonus: the product of all triples of
operators along the lines gives $III$, again up to a sign. In this
way we have obtained a sort of {\it non}-commutative labelling for
the points of GQ$(2,4)$. The $15$ points of the doily are labelled
by $15$ symmetric operators, and the $12$ double-sixes are
labelled by antisymmetric ones. The incidence relation on this set
of $27$ points producing the $45$ lines is: a pairwise commuting
property and a ``sum rule" (i.\,e. multiplication producing $III$
up to a sign).

Notice that for $a,b,c=1,2,\dots,7$ the combinations $g_{abc}\equiv g_ag_bg_c$ as elements of the Clifford algebra $Cliff(7)$ are
symmetric and the ones $g_a$ and $g_{ab}=g_ag_b$
are antisymmetric matrices with some signs automatically incorporated.
Hence, the simplest choice for a labelling taking care of the signs  is simply
\beq
\{1^{\prime},2^{\prime},3^{\prime},4^{\prime},5^{\prime},6^{\prime}\}=
\{g_2,g_3,g_4,g_5,g_6,g_7\},
\label{1v}
\eeq
\noindent
\beq
\{1,2,3,4,5,6\}=\{g_{12},g_{13},g_{14},g_{15},g_{16},g_{17}\},
\label{2v}
\eeq
\noindent
\beq
\{12,13,14,15,16,23,24,25,26\}=
\{g_{123},g_{124},g_{125},g_{126},g_{127},g_{134},g_{135},g_{136},g_{137}\},
\label{3v}
\eeq
\noindent
\beq
\{34,35,36,45,46,56\}=
\{g_{145},g_{146},g_{147},g_{156},g_{157},g_{167}\}.
\label{4v}
\eeq
\noindent
Using the explicit form of the $8 \times 8$ matrices $g_a, a=1,2,\dots,7$ of Eq.\,(\ref{cliff}), we can get three-qubit operators with
a natural choice of signs as non-commutative labels for the points of GQ$(2,4)$.
The summary of this chain of reasoning is displayed in Figure 3.

\begin{figure}[t]
\centerline{\includegraphics[width=10truecm,clip=]{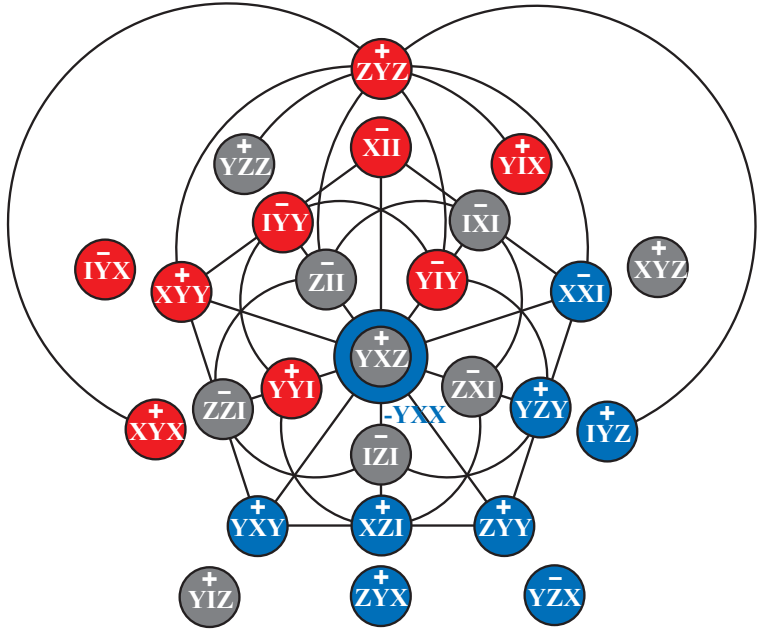}}
\vspace*{0.5cm} \caption{An illustration of the non-commutative labelling of the points of GQ$(2,4)$. For better readability of the figure,
the sign of an operator is placed above the latter.}
\end{figure}
\subsection{The Weyl action on GQ$(2,4)$}

Using our new labelling we can demonstrate the $W(E_6)$ invariance of GQ$(2,4)$
This renders our arguments on the relationship between the structure of GQ$(2,4)$ and $I_3$ to a proof.

Let us consider the correspondence
\beq
I\mapsto (00),\qquad X\mapsto (01),\qquad Y\mapsto (11),\qquad Z\mapsto (10).
\eeq
\noindent
Using this we can map an arbitrary element  of the central quotient of the three-qubit Pauli group to ${\bf Z}_2^6$, i.\,e. to the space of $6$-component vectors with elements taken from $GF(2)$.
For example, $XZI$ is taken to the $6$-component vector $(011000)$.
Clearly, if we are interested merely in the incidence structure then we can label the points of GQ$(2,4)$ with such six component vectors.
Knowing that $W(E_6)\cong O^-(6,2)$, which is the set of $6\times 6$ matrices with entries taken from $GF(2)$ leaving invariant a special quadratic form \cite{Shaw} defined on 
${\bf Z}_2^6$, we can check the Weyl invariance by checking  the invariance under a suitable set of generators.
From the atlas of finite groups \cite{atlas} we use the presentation
\beq
O^-(6,2)=U(4,2)\rtimes {\bf Z}_2=\langle c,d\vert c^2=d^9=(cd^2)^8=[c,d^2]^2=[c,d^3cd^3]=1\rangle.
\eeq
\noindent
We have found the following representation convenient (this is the one that is preserving the symplectic structure corresponding to the commutation properties in the Pauli group \cite{Nielsen}, and mapping the $27$ three-qubit label onto itself)
\beq
c=\begin{pmatrix}1&0&0&0&0&0\\1&1&1&1&0&0\\1&0&0&1&0&0\\
1&0&1&0&0&0\\0&0&0&0&1&0\\0&0&0&0&0&1
\end{pmatrix},\quad
d=\begin{pmatrix}0&1&1&1&1&1\\1&1&0&1&0&1\\1&1&1&0&1&1\\0&0&1&1&1&1\\0&1&1&1&0&0\\0&1&1&0&0&1\end{pmatrix}.
\eeq
\noindent
Using the above-given dictionary, we explicitly get
for the action of $c$
\beq
IXI\mapsto XZI,\qquad ZYX\mapsto YIX,\qquad IZI\mapsto XXI
\eeq
\noindent
\beq
ZYZ\mapsto YIZ,\qquad ZII\mapsto YYI,\qquad ZYY\mapsto YIY,
\eeq
\noindent
and the remaining $15$ operators are left invariant.
For the action of $d$ we get three orbits
\beq
IXI\mapsto YXZ\mapsto YZX\mapsto YIX\mapsto XYZ\mapsto IYZ\mapsto YXX\mapsto ZZI\mapsto YXY
\eeq
\noindent
\beq
IZI\mapsto ZYY\mapsto XII\mapsto YZY\mapsto XYX\mapsto XYY\mapsto YIY\mapsto YIZ
\mapsto IYY
\eeq
\noindent
\beq
IYX\mapsto ZXI\mapsto ZYZ\mapsto ZYX\mapsto YYI\mapsto YZZ\mapsto ZII\mapsto XZI\mapsto XXI.
\eeq
\noindent
One can check that these generators take lines to lines, hence preserving GQ$(2,4)$.

Moreover, using this action of $W(E_6)$ on GQ$(2,4)$ we can define a corresponding one on GQ$(2,4)$ taken together with the non-commutative coordinates.
For this we can take the very same expressions as above but taking also into account the {\it signs} of the operators, as shown in Figure 3.
Since these signed quantities automatically take care of the structure of signs of $I_3$, this furnishes a proof for the $W(E_6)$ invariance of $I_3$.
Notice also that the transformation rules for the non-commutative labels
imply the corresponding rule for the {\it charges}.
Having in this way an explicit action on the charges and the invariance of the black hole entropy, it would be interesting to work out manifestations of 
this discrete symmetry of order $51840$ in string theory.

\subsection{A $D=4$ interpretation}

Note that
the decomposition
\beq
E_{7(7)}\supset E_{6(6)}\times SO(1,1)
\label{45decomp}
\eeq
\noindent
under which
\beq
{\bf 56}\rightarrow {\bf 1}\oplus{\bf 27}\oplus{\bf 27}^{\prime}\oplus{\bf 1}^{\prime}
\eeq
\noindent
describes the relation between the $D=4$ and $D=5$ duality groups \cite{FKdual,KK,Laura3}.
We intend to show that the non-commutative labelling constructed for our quadrangle provides a nice finite geometric interpretation
of the physics based on
the decomposition of Eq.\,(\ref{45decomp}).

To this end, we use the $N=8$ central charge parametrized as in Eq.\,(\ref{central4}) and look at the structure of the cubic invariant
that can be written also in the alternative form \cite{Becker}
\beq
I_3=\frac{1}{48}{\rm Tr}(\Omega {\cal Z}\Omega {\cal Z}\Omega {\cal Z})
\label{nicecube}
\eeq
\noindent
where for $\Omega$ we use the definition of Eq.\,(\ref{omega}).
In order to get the correct number of components, we impose the usual constraints \cite{Gimon}
\beq
{\rm Tr}(\Omega {\cal Z})=0,\qquad \overline{\cal Z}=\Omega {\cal Z}{\Omega}^T.
\label{cubeconstr}
\eeq
\noindent
Notice that the first of these constraints restricts the number of antisymmetric matrices to be considered in the expansion of
${\cal Z}$ from $28$ to $27$.
The second constraint is the usual reality condition which restricts the $27$
{\it complex} expansion coefficients to $27$ {\it real} ones, producing the right count.
Recall also that the group theoretical meaning of these constraints is the expansion
of the $N=8$ central charge in an $USp(8)$ basis, which is appropriate since
$USp(8)$ is the automorphism group of the $N=8$, $D=5$ supersymmetry algebra.

It is easy to see that the reality constraint yields
\beq
y_{jk}=0,\qquad x^{0j}=0,\qquad x^{1j}=0,\qquad j,k=2,3,\dots\,7,
\eeq
\noindent
hence $\Omega{\cal Z}$ is of the form
\beq
\Omega{\cal Z}={\cal S}+i{\cal A}\equiv \frac{1}{2}x^{jk}g_{1jk}+i(y_{0j}g_{1j}-y_{1j}g_j),
\eeq
\noindent
where summation for $j,k=2,3,\dots,7$ is understood.
The new notation for ${\Omega}{\cal Z}$ shows that ${\cal S}$ is {\it symmetric} and ${\cal A}$ is {\it antisymmetric}.
Notice that the three-qubit operators occurring in the expansions of ${\cal S}$
and ${\cal A}$ are precisely the ones we used in Eqs. (\ref{1v})--(\ref{4v})
as our non-commutative ``coordinates" for GQ$(2,4)$.

Performing standard manipulations, we get
\beq
I_3=\frac{1}{48}\left( {\rm Tr}({\cal SSS})-3{\rm Tr}({\cal SAA})\right).
\label{i3vv}
\eeq
\noindent
                                                                                Notice that                                                                     \beq                                                                            \Omega g_{i+1} g_{j+1} g_{k+1} g_{l+1} g_{m+1} g_{n+1}=-{\varepsilon}_{ijklmn},   \quad i,j,k,l,m,n=1,2,\dots,6,\quad {\varepsilon}_{123456}=+1.                     \label{epsilon}                                                                 \eeq                                                                            \noindent
Hence, with the notation
\beq
A^{jk}\equiv x^{j+1k+1}, \qquad u_j\equiv y_{0j+1},\qquad v_j\equiv y_{1j+1},\qquad j,k=1,2,\dots,6,
\eeq
\noindent
the terms of Eq.\,(\ref{i3vv}) give rise to the form of Eq.\,(\ref{Pfaffdecomp}).
Also notice that the parametrization
\beq
u^T=\begin{pmatrix}b^{10},&-c_{10},&{a^1}_2,&c_{12},&b^{12},&{a^1}_0            \end{pmatrix},
\eeq
\noindent
\beq
v^T=\begin{pmatrix}-c_{21},&-{a^{2}}_1,&-b^{01},&-{a^0}_1,&c_{01},&b^{21}       \end{pmatrix},
\eeq
\noindent
\beq
A=\begin{pmatrix}0&c_{02}&b^{22}&-c_{00}&{a^1}_1&b^{02}\\
-c_{02}&0&{a^0}_0&b^{11}&c_{22}&{-a^0}_2\\
-b^{22}&-{a^0}_0&0&{a^2}_0&b^{20}&c_{11}\\
c_{00}&-b^{11}&-{a^2}_0&0&c_{20}&{a^2}_2\\
-{a^1}_1&-c_{22}&-b^{20}&-c_{20}&0&-b^{00}\\
-b^{02}&{a^0}_2&-c_{11}&-{a^2}_2&b^{00}&0\end{pmatrix},
\eeq
\noindent
yields the qutrit version of $I_3$ of Eq.\,(\ref{i3v}).

The main message of these considerations is obvious: different versions
of $I_3$, and, so, of the black hole entropy formula, are obtained as different
parametrizations of the underlying finite geometric object --- our generalized quadrangle GQ$(2,4)$.

\subsection{Mermin squares and the hyperplanes of the hexagon}

Our non-commutative coordinatization of GQ$(2,4)$ in terms of the elements of $Cliff(7)$, or equivalently by three-qubit operators, is very
instructive. For example, one can easily check that this labelling for the doily
gives rise to $7$ lines with a minus sign and $8$ lines with a plus one.
(That is, the product of the corresponding operators yields either $-III$ or $+III$.)
This is in accord with the sign structure of the Pfaffian. It is easy to check that for each of the $10$ grids living inside the doily
these signs give rise to $3$ plus signs and $3$ minus ones needed for producing the determinant related to the $10$ possible truncations
with $9$ charges. These $10$ grids generate $10$ Mermin squares.

As repeatedly mentioned, inside GQ$(2,4)$ there are also triads of grids which are partitioning its $27$ points.
These are the ones related to the three qutrit states, indicated by coloring the corresponding points in three different ways.
They are also producing Mermin squares.
Note, the usual definition of a Mermin square is a grid having the property
that the products of operators along any of its rows and columns {\it except for one} yield $III$. Here, we define Mermin squares as
objects for which no simple sign assignment can produce the rule the operator products give.
In this generalized sense we have $3\times 40=120 $ Mermin squares living inside our GQ$(2,4)$.

Of course, our particular ``coordinates" producing the three special Mermin-squares
can be replaced by other possible ones arising from $27$ further labellings.
In order to see this notice that the``non-commutative coordinates" of Figure 3
are the ones based on a special choice for the matrix ${\Omega}$ of Eq.\,(\ref{omega}).
Since we have $28$ antisymmetric operators, we have $27$ further possible choices
for ${\Omega}$.
Choosing {\it any} of these matrices will produce a $27=12+15$ split for the space of the remaining antisymmetric operators.
For this we simply have to consider the $12$ operators {\it anticommuting} and the $15$ ones {\it commuting} with our fixed
${\Omega}$. This can be easily checked using the property that the antisymmetric matrices are either of the form $g_{ab}$ or
$g_a , a=1,2,\dots,7$.
Now apply the simple rule: multiply the $15$ operators commuting with ${\Omega}$ by ${\Omega}$ and leave the remaining ones untouched.
One can then check that this procedure will yield $27$ further possible non-commutative labels for the points of GQ$(2,4)$,
hence another possible sets of Mermin squares.
Notice also that for the special choice ${\Omega}=YYY$ the reality condition of Eq.\,(\ref{cubeconstr}) is related to the three-qubit
version of the so-called  Wootters spin-flip operation \cite{Wootters} used in quantum information.

We round off this section with an important observation.
The Coxeter set comprising the points of the generalized hexagon of order two
answering to the set of antisymmetric three-qubit operators is just the complement of one of the geometric hyperplanes of the hexagon.
The $28$ possibilities for fixing ${\Omega}$ gives rise to $28$ subconfigurations consisting of $27$ points.
These $27$ points are always consisting of $12$ antisymmetric operators and $15$ symmetric ones.
By picturing them inside the hexagon \cite{Lev4}, one can realize that any such subconfiguration consists of $9$ pairwise disjoint
lines (i.\,e., is a distance-3-spread).
It turns out that these subconfigurations are also {\it geometric hyperplanes}
living inside the hexagon \cite{Frohardt}.
Hence, we have found a very interesting geometric link between the structures of $D=4$ and $D=5$ entropy formulas.
The $D=4$ case is related to the split Cayley hexagon of order two \cite{Lev4} and here we have demonstrated that the $D=5$ one
is underlined by the geometry of the generalized quadrangle GQ$(2,4)$.
The connection between these cases is based on a beautiful relationship
between the structure of GQ$(2,4)$ and one of the geometric hyperplanes of the hexagon.

\section{Conclusion}
In this paper we revealed an intimate connection between the structure of black hole entropy formulas in $D=4$ and $D=5$ and the geometry
of certain finite generalized polygons.
We provided a detailed correspondence between the structure of the cubic invariant related to the black hole entropy in $D=5$ and the
geometry of the generalized quadrangle GQ$(2,4)$ with automorphism group the Weyl group $W(E_6)$. In this picture the $27$ charges
correspond to the points and the
$45$ terms in the entropy formula to the lines of GQ$(2,4)$. Different truncations with $15, 11$ and $9$ charges are represented by three
distinguished subconfigurations of GQ$(2,4)$, well-known to finite geometers; these are the ``doily" (i.\,e. GQ$(2,2)$) with $15$, the
``perp-set" of a point with $11$, and the ``grid" (i.\,e. GQ$(2,1)$) with $9$ points, respectively.
Different truncations naturally employ objects like cubic Jordan algebras well-known to string theorists,  or qubits and qutrits
well-known to quantum information theorists.
In our finite geometric treatment these objects just provide different
coordinates for the underlying geometric object, GQ$(2,4)$.
However, in order to account also for the signs of the monomials in the qubic invariant, the labels, or ``coordinates"
used for the points of GQ$(2,4)$ must be non-commutative.
We have shown that the real operators of the three-qubit Pauli group
provide a natural set of such coordinates.
An alternative way of looking at these coordinates is obtained by employing a special $27$ element set of $Cliff(7)$.
Hence it seems quite natural
to
conjecture that the different  possibilities of
 describing the $D=5$ entropy formula using Jordan algebras, qubits and/or qutrits merely correspond  to employing different coordinates for an
 underlying noncommutative geometric structure based on GQ$(2,4)$.

Using these coordinates we established the Weyl invariance of the cubic invariant and we also shed some light on the interesting connection between the different possible truncations with $9$
charges and the geometry of  Mermin squares  --- objects well-known from studies
concerning Bell-Kochen-Specker like theorems.
Since these $9$-charge configurations as qutrits can also be connected to special brane configurations \cite{Duff2},
it would also be nice to relate their physical properties to these Mermin squares.

We emphasize that these results are also connected to our previous ones obtained for the $E_7$ symmetric entropy formula in $D=4$
by observing that the structure of GQ$(2,4)$ is linked to a particular geometric hyperplane of the split Cayley hexagon of order two
\cite{Lev4} featuring $27$ points located on $9$ pairwise disjoint lines (a distance-3-spread).
This observation provides a direct finite geometric link between the $D=4$ and $D=5$ cases.
However, there are other interesting hyperplanes of the hexagon.
Their physical meaning (if any) is not clear.
In particular, we have other three distinct types of hyperplanes with $27$ points inside the hexagon \cite{Frohardt}.
They might shed some light on the geometry of further truncations
that are not arising so naturally as the ones discussed in this paper.

Finally, it is worth mentioning that the above-employed generalized quadrangles with lines of size three are also closely related with particular root lattices \cite{brouwer-haemers}. Given an irreducible root lattice $\Lambda$, one picks any two roots $a, b$ whose inner product equals unity, $\langle a, b \rangle = 1$ (whence $\langle a, a \rangle =  \langle b, b \rangle = 2$). Then the set $S = \{r \in \Lambda |  \langle r, r \rangle = 2, \langle r, a \rangle = \langle r, b \rangle = 1 \}$ is a generalized quadrangle with lines of size three  if the latter are represented by the triples $\{x,  y,  z\}$ meeting the constraint $x + y + z = a + b$. Since $\Lambda$ is spanned by $\{ a, b \} \cup S$, the structure of $S$ determines $\Lambda$. And it turns out \cite{brouwer-haemers} that the root lattices that correspond to GQ$(2, 1)$, GQ$(2, 2)$, and GQ$(2, 4)$, are nothing but those of $E_6$, $E_7$ and $E_8$, respectively.

\section*{Acknowledgement}
This work was partially supported by the VEGA grant agency projects Nos. 2/0092/09 and 2/7012/27.

\end{document}